%% file: main.tex
\pdfoutput=1

\documentclass[sigplan,10pt]{acmart}
\usepackage{extarrows}

\usepackage{tabularx}
\usepackage{booktabs}
\usepackage{multicol}
\usepackage{multirow}
\usepackage{threeparttable}

\usepackage{amsmath,amssymb,amsfonts,mathtools}
\usepackage{algorithmic}
\usepackage{graphicx}
\usepackage{textcomp}
\usepackage{xcolor}
\def\BibTeX{{\rm B\kern-.05em{\sc i\kern-.025em b}\kern-.08em
		T\kern-.1667em\lower.7ex\hbox{E}\kern-.125emX}}

\usepackage{caption}
\usepackage{subcaption}
\usepackage{balance} 
\usepackage{xspace}
\usepackage[T1]{fontenc}
\usepackage[scaled=0.81]{beramono}
\usepackage{balance}

\usepackage{enumitem}  
\usepackage{listings}
\usepackage{url}
\usepackage[ruled,lined,linesnumbered,vlined]{algorithm2e}
\usepackage{multirow}
\usepackage{color}

\input{solidity-highlighting}
\definecolor{darkgreen}{rgb}{0.0, 0.5, 0.13}

\newcommand{\tool}{\hbox{\textsc{TCWasm}}\xspace}
\newcommand{\ethereum}{\hbox{Ethereum}\xspace}
\newcommand{\tcpa}{\hbox{TCPA}\xspace}
\newcommand{\tee}{\hbox{TEE}\xspace}

\newcommand{\cfg}{\hbox{CFG}\xspace}

\newcommand{\wasm}{\hbox{WASM}\xspace}

\newcommand{\neval}{\hbox{33}\xspace}

\newcommand{\overheadtime}{\hbox{139.3\%}\xspace}
\newcommand{\overheadmemory}{\hbox{54.8\%}\xspace}

\newcommand{\myparagraph}[1]{\vspace*{0.14cm}\noindent\textbf{\emph{#1.}}\quad}

\newcommand{\etal}{\hbox{\emph{et al.}}\xspace}
\newcommand{\eg}{\hbox{\emph{e.g.}}\xspace}
\newcommand{\ie}{\hbox{\emph{i.e.}}\xspace}

\newcommand{\etc}{\hbox{\emph{etc.}}\xspace}

\newcommand{\dataset}{\texttt{\url{http://omitted.for.double-blind-review/}}}

\newcommand{\RoT}{\textit{RoT}}
\newcommand{\Plat}{\textit{Plat}}
\newcommand{\IC}{\textit{IC}}
\newcommand{\pb}{\textit{pub}}
\newcommand{\pv}{\textit{priv}}
\newcommand{\TCPAIC}{\textit{TCPA-IC}}
\newcommand{\PC}{\textit{PC}}
\newcommand{\ICC}{\textit{ICC}}
\newcommand{\CC}{\textit{CC}}
\newcommand{\OC}{\textit{OC}}

\AtBeginDocument{%
\providecommand\BibTeX{{%
		\normalfont B\kern-0.5em{\scshape i\kern-0.25em b}\kern-0.8em\TeX}}}

\begin{document}

\title{Trusted And Confidential Program Analysis}


\author{Han Liu}
\affiliation{%
	\institution{The Blockhouse Technology Ltd.}
	\state{Oxford}
	\country{UK}
}
\email{liuhan@tbtl.com}

\author{Pedro Antonino}
\affiliation{%
	\institution{The Blockhouse Technology Ltd.}
	\state{Oxford}
	\country{UK}
}
\email{pedro@tbtl.com}

\author{Zhiqiang Yang}
\affiliation{%
	\institution{The Blockhouse Technology Ltd.}
	\state{Oxford}
	\country{UK}
}
\email{zhiqiang@tbtl.com}

\author{Chao Liu}
\affiliation{%
	\institution{The Blockhouse Technology Ltd.}
	\state{Oxford}
	\country{UK}
}
\email{liuchao@tbtl.com}

\author{A.W. Roscoe}
\affiliation{%
	\institution{The Blockhouse Technology Ltd.}
	\institution{University College Oxford Blockchain Research Center}
	\institution{Oxford University}
	\state{Oxford}
	\country{UK}
}
\email{awroscoe@gmail.com}

\input{abstract}

\section{Introduction}
\label{sec:intro}
\input{intro}

\section{Background}
\label{sec:bg}
\input{bg}

\section{Protocol For \tcpa}
\label{sec:method}
\input{method}

\section{Design of \tool}
\label{sec:design}
\input{design}

\section{Preliminary Evaluation}
\label{sec:eval}
\input{eval}

\section{Related Work}
\label{sec:rw}
\input{rw}

\section{Conclusion}
\label{sec:conclusion}
\input{conclusion}


\bibliographystyle{ACM-Reference-Format}
\bibliography{seraph,tee,program,blockchain}

\end{document}

%% file: solidity-highlighting.tex
\usepackage{listings, xcolor}

\definecolor{verylightgray}{rgb}{.97,.97,.97}

\lstdefinelanguage{Solidity}{
	keywords=[1]{anonymous, assembly, assert, balance, break, call, callcode, case, catch, class, constant, continue, contract, debugger, default, delegatecall, delete, do, else, emit, event, export, external, false, finally, for, function, gas, if, implements, import, in, indexed, instanceof, interface, internal, is, length, library, log0, log1, log2, log3, log4, memory, modifier, new, payable, pragma, private, protected, public, pure, push, require, return, returns, revert, selfdestruct, send, storage, struct, suicide, super, switch, then, this, throw, transfer, true, try, typeof, using, value, view, while, with, addmod, ecrecover, keccak256, mulmod, ripemd160, sha256, sha3}, 
	keywordstyle=[1]\color{blue}\bfseries,
	keywords=[2]{address, bool, byte, bytes, bytes1, bytes2, bytes3, bytes4, bytes5, bytes6, bytes7, bytes8, bytes9, bytes10, bytes11, bytes12, bytes13, bytes14, bytes15, bytes16, bytes17, bytes18, bytes19, bytes20, bytes21, bytes22, bytes23, bytes24, bytes25, bytes26, bytes27, bytes28, bytes29, bytes30, bytes31, bytes32, enum, int, int8, int16, int24, int32, int40, int48, int56, int64, int72, int80, int88, int96, int104, int112, int120, int128, int136, int144, int152, int160, int168, int176, int184, int192, int200, int208, int216, int224, int232, int240, int248, int256, mapping, string, uint, uint8, uint16, uint24, uint32, uint40, uint48, uint56, uint64, uint72, uint80, uint88, uint96, uint104, uint112, uint120, uint128, uint136, uint144, uint152, uint160, uint168, uint176, uint184, uint192, uint200, uint208, uint216, uint224, uint232, uint240, uint248, uint256, var, void, ether, finney, szabo, wei, days, hours, minutes, seconds, weeks, years},	
	keywordstyle=[2]\color{teal}\bfseries,
	keywords=[3]{block, blockhash, coinbase, difficulty, gaslimit, number, timestamp, msg, data, gas, sender, sig, value, now, tx, gasprice, origin},	
	keywordstyle=[3]\color{violet}\bfseries,
	identifierstyle=\color{black},
	sensitive=false,
	comment=[l]{//},
	morecomment=[s]{/*}{*/},
	commentstyle=\color{gray}\ttfamily,
	stringstyle=\color{red}\ttfamily,
	morestring=[b]',
	morestring=[b]"
}

%% file: abstract.tex
\begin{abstract}

We develop the concept of Trusted and Confidential Program Analysis (\tcpa) which enables 
program certification to be used where previously there was insufficient trust. 
Imagine a scenario where a producer may not be trusted to certify its own software (perhaps by a foreign regulator), 
and the producer is unwilling to release its sources and detailed design to any external body. 
We present a protocol that can, using trusted computing based on encrypted sources, create 
certification via which all can trust the delivered object code without revealing the 
unencrypted sources to any party. Furthermore, we describe 
a realization of \tcpa with trusted execution environments (\tee) that enables general and efficient 
computation. We have implemented the \tcpa protocol in a system called \tool for web assembly 
architectures. In our evaluation with \neval benchmark cases, \tool managed to finish the 
analysis with relatively slight overheads.


\end{abstract}

\ccsdesc[300]{Security and privacy~Domain-specific security and privacy architectures}

\keywords{Program analysis, regulatory property, trusted execution environment}



\maketitle

%% file: intro.tex
\begin{figure}[h]
\centering
\includegraphics[width=.8\linewidth]{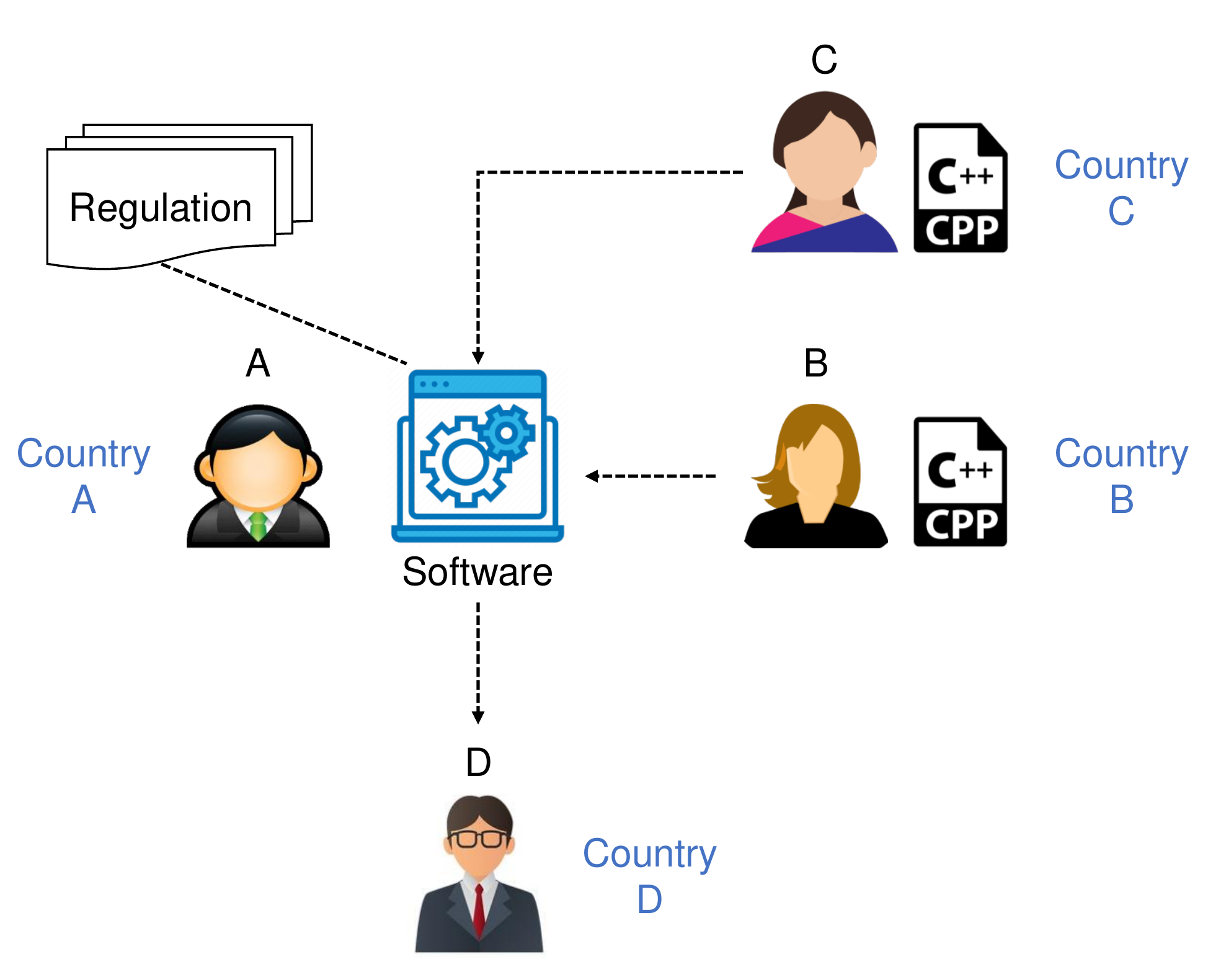}
\caption{\label{fig:illustrative}Multiparty trusted and confidential regulation.}
\end{figure}

This paper offers a tentative solution to the following problem: vendor Alice wishes to sell executable software E to clients such as Bob. Bob does not trust Alice and wants to examine the high-level design or sources behind E, but Alice is unwilling to do this to protect her intellectual property. 
In particular, the rapid growth of the international IT market, where IT products (\eg, software, hardware, services, \etc) 
are widely exported from one country to another has led to many such issues. Unfortunately, IT 
products are notoriously prone to bugs and security risks no matter in what 
forms they are deployed and in what environments they are running. Any vulnerable component 
in an IT product may cause a great loss to customers, \eg, national security 
threats, abuse of personal data, manipulation on digital asset, \etc In practice, the 
import and export of IT products are commonly required to be strictly compliant with 
local and international regulations. More importantly, such regulatory processes have to 
take place in a both \emph{trusted} and \emph{confidential} way, \ie, form a consensus 
on the regulatory compliance of an IT product among a group of relevant parties without 
leaking further sensitive information, \eg, source code, hardcoded values, \etc Figure~\ref{fig:illustrative} 
shows an illustrative example for further explanation.

\myparagraph{Illustrative Example}
We consider the case in which \emph{B} and \emph{C} have jointly developed software and exported it to \emph{A}. 
As a service provider, \emph{A} imported the software to set up a public service which further reached 
\emph{D} as a general user. Although this case is designed for explanation, it actually 
abstracts a typical scenario where nowadays IT products are deployed and used worldwide. 
Considering that the four participants may be from different countries, 
they are required to follow various regulations. 
For \emph{B} and \emph{C} as software exporters, they need to be compliant with local export regulations. 
For \emph{A} as a technology importer, his or her obligation is to check that the imported software introduces no 
regulatory risks, \eg, national security issues. From the perspective of \emph{D}, his or her country might 
only allow the service to reach domestic users if it poses no threats to the public good, \eg, user 
privacy. In practice, it is hard to enforce the regulatory processes among these 
four parties due to the fact that the source code of the software is strictly confidential 
and therefore cannot be directly shared in a straightforward manner. Consequently, 
no single party in this case is able to believe that the software indeed delivers the 
required regulatory compliance across different countries.

\myparagraph{Trusted and Confidential Program Analysis}
To address this problem, we proposed a novel protocol in this paper to enable trusted and confidential 
program analysis (\tcpa) for checking regulatory compliance of software across multiple parties without mutual 
trust. More specifically, the proposed protocol guarantees that a) imported or exported software $E$ 
is indeed built from a given piece of secret source code $C$, b) both $E$ and $C$ are compliant with a 
set of mutually agreed regulatory properties $P=\{ p_1, p_2, \cdots, p_n \}$, \ie, $E, S \vDash P$ and 
c) the compliance of $E$ and $S$ is verifiable without revealing sensitive information of $S$. 
Furthermore, we described a realization of \tcpa called \tool using 
trusted execution environments (\tee{}s) and applied the system for analysis of web assembly (\wasm) programs. In the preliminary evaluation with \neval 
benchmark files, \tool finished the analysis tasks with slight overheads of \overheadtime and 
\overheadmemory in terms of time and memory usage, respectively; the baseline system executes the same analysis tasks without relying on TEEs. These overheads seem entirely acceptable to us given the added guarantees in terms of trust and confidentiality that the use of TEEs provides.

We summarize our main contributions below.
\begin{itemize}[leftmargin=*]
\item We describe the problem of trusted and confidential program analysis and present a 
formalization of it to guide our subsequent research.

\item We propose the first protocol to enable \tcpa in practice, which is consistent with 
popular trusted computing technologies such as \tee{}s. 

\item We realize the protocol using AMD's SEV \tee implementation. We develop the system \tool for verifying web assembly programs 
via \tcpa; the first of its kind, to the best of the authors' knowledge.

\item We have conducted a large-scale evaluation of \tool and here report empirical evidence to 
demonstrate the feasibility of applying \tcpa in practice.
\end{itemize}

\myparagraph{Paper Organization}
The rest of this paper is organized as follows. \S\ref{sec:bg} introduces background information. 
\S\ref{sec:method} presents an in-depth explanation of the \tcpa protocol. \S\ref{sec:design} 
describes the system design of \tool. \S\ref{sec:eval} summarizes empirical results of the 
evaluation and \S\ref{sec:rw} discusses related works. \S\ref{sec:conclusion} concludes the paper.


%% file: bg.tex
\subsection{Trusted Computing}
\label{subsec:tc}
\emph{Trusted computing}~\cite{mitchell2005trusted, tc2013trusted} describes a number of technologies proposed to 
mitigate security threats. Unlike conventional passive approaches such as firewalls, 
malware detection, and intrusion detection, trusted computing takes a more active approach to 
addressing these threats, \ie, by establishing trust on a solid basis (typically referred to as the \emph{root of trust} and implemented using both 
hardware and software) and then expanding this chain of trust to cover the entire 
computing system. Trusted Execution Environments represent arguably the most flexible and practical technology in this area. Some modern processors offer hardware primitives that support the execution of \emph{isolated computations}, that is, the code and data they use are encrypted and integrity-protected. These TEE implementations are designed so that even the operator of the machine where such an execution takes place is unable to undetectably tamper with, or learn secrets manipulated by this execution. We use the terms TEE and TEE implementation interchangeably. Note that a TEE can be managing and executing multiple isolated computations concurrently. Examples of TEE implementations available are
Intel’s Software Guard Extensions (SGX)~\cite{costan2016intel}, 
AMD’s Secure Encrypted Virtualization (SEV)~\cite{sev2020strengthening}, and 
ARM’s TrustZone~\cite{pinto2019demystifying}.  Each of them is designed to address different application scenarios, but they all share similar core capabilities. We introduce some of these common building blocks as follows. 

These implementations rely on special instructions to load the isolated computation's code and data, in encrypted form, into the \emph{main memory}, while \emph{measuring} them in the process, and execute it. They isolate computations at different levels of granularity. For instance, while AMD SEV isolates an entire virtual machine (VM), Intel SGX isolates part of a (operating-system) process. This measurement and the encrypted memory pages are created, managed, and protected by the TEE. The \emph{attestation} process plays a key part in establishing the chain of trust. Roughly speaking, it generates cryptographically-protected evidence attesting that a computation with a given measurement has been properly isolated; it also attests the authenticity of the TEE implementation - hardware and software components - and its capabilities. By properly verifying this evidence, an entity interacting with this computation can be confident of its isolation and, consequently, of the guarantees that follow. We detail this process later. For instance, in the case of AMD SEV, the attestation process can be used to establish that a VM has been properly set up by checking the boot process and the elements it depends upon have the expected measurements, i.e. OS image, installed programs and data, etc. The isolation offered by TEE implementations is restricted to the main memory. The developer of the isolated computation is responsible for encrypting data on disk.

\subsection{Program Analysis}
\label{subsec:pa}
In general, program analysis develops on methods and technologies at establishing the relationship 
between a given software program and expected properties, \eg, correctness, robustness, safety 
and liveness. Typical program analysis tasks are program optimization, vulnerability detection, 
and formal verification. Considering the approaches taken, program analysis technologies can be 
categorized as static program analysis, dynamic program analysis, and both in combination.
Static program analysis technology finishes its task by not running or executing the target program. 
Common static program analysis methods are control-flow analysis, data-flow analysis, 
abstract interpretation, type analysis, and pointer analysis. All of these methods first model 
the given program and its expected property abstractly as a set of mathematical constraints, for which the analysis task boils down to establishing satisfiability. 
In contrast, dynamic program analysis fulfills its goal by examining actual runs of the target program  and checking for the expected properties. 
Fuzzing, symbolic execution, and runtime monitoring are among the
most frequently used dynamic program analysis technologies. While static and dynamic program analysis 
have their strengths in efficiency and precision, respectively, neither of them is able to tackle all 
real-world problems. Thus, there is another category of \emph{hybrid} program analysis technologies which 
combines both approaches. \tcpa encompasses all of these.

\begin{figure*}[h]
	\centering
	\includegraphics[width=.8\linewidth]{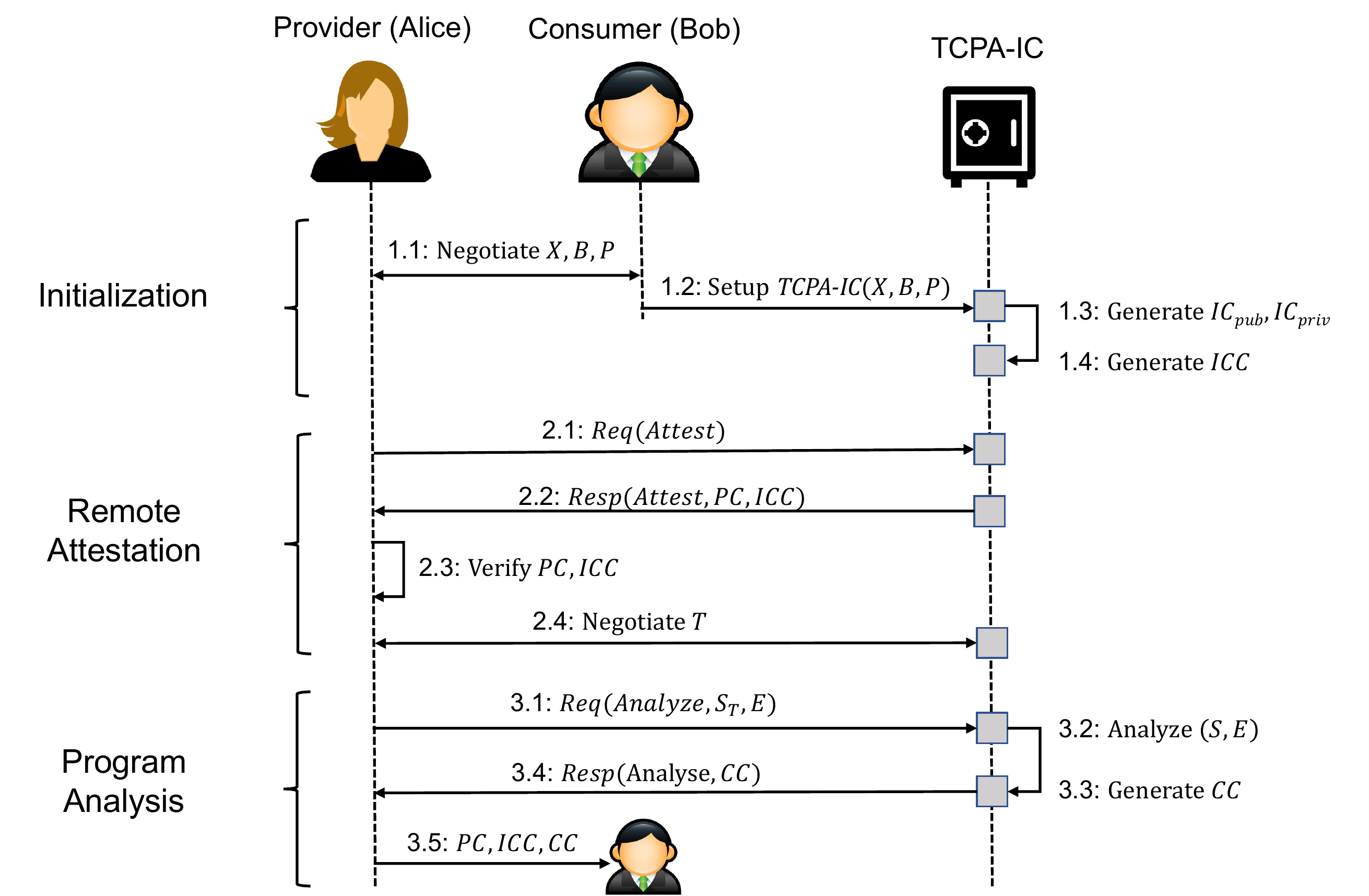}
	\caption{\label{fig:protocol}The protocol of \tcpa.}
\end{figure*}

%% file: method.tex
\subsection{Overview}
\label{subsec:overview}

The general workflow of \tcpa is shown in Figure~\ref{fig:protocol}. The protocol defines the following types of actors, namely, Provider (Alice), Consumer (Bob), and the Trusted and Confidential Program Analysis Isolated Computation (\TCPAIC{}). Alice is the software provider in this 
setting who wants to export her products to the Consumer Bob and therefore needs to convince 
him that the exported software is strictly compliant with regulations. 
As aforementioned, Alice also needs to avoid revealing the product's business secrets 
(\eg, source code) to non-relevant parties such as Bob. In this set-up, Bob hosts the \tee{} where \TCPAIC{} is to be executed. He is in charge of setting up the TEE-enabled platform and \TCPAIC{}. Even with unlimited access to the platform, not even Bob can infer any secrets from this compliance analysis since the product is analysed within the isolated service \TCPAIC{}. Of course, we could have an alternative setup where Alice or a third party would host the \tee where \TCPAIC{} executes. This set-up, however, reinforces and highlights that even the platform's operator, with unrestricted access, cannot read into the isolated computation's execution.

Important concepts in \tcpa as listed below.

\begin{itemize}[leftmargin=*]
\item $S$ is the source code of a given piece of software.

\item $E$ is the executable corresponding to $S$.

\item $P=\{p_1, p_2, \cdots, p_n\}$ is a set of regulatory properties to be checked.

\item $X$ is a program analysis framework, \eg, static analysis, dynamic analysis. 

\item $B$ is the building framework of $S$, \ie, create $E$ from $S$. The most common example of 
$B$ is a compiler.

\item $\TCPAIC{}(X,B,P)$ is the isolated computation that carries out the indicated trusted and confidential program analysis.

\item Public and private keys \IC{}$_{\pb{}}$ and \IC{}$_{\pv{}}$, respectively, are elements of a signature scheme used to authenticate messages issued by $\TCPAIC{}(X,B,P)$. 

\item $T$ is a symmetric key for encryption and decryption. $S_T$ indicates an encrypted version of $S$.

\item The platform certificate (\PC{}), isolated computation certificate (\ICC{}), and compliance certificate (\CC{}) attest the platform's capabilities, the isolated computation's measurement, and the program analysis' outcome, respectively - we detail these elements in the following.

%
%
%
%

\end{itemize}

\subsection{Platform set-up}

Before the protocol in Figure~\ref{fig:protocol} can be executed, we assume a \emph{platform setup} step; omitted from that figure for the sake of simplicity. In this step, the platform engage in a protocol with the processor manufacturer, who is typically also the TEE implementer, aimed at proving the processor's authenticity, its capability to run isolated computations, and that all the trusted elements - both hardware and software - part of the TEE implementation are valid and have been properly set up. Manufacturers store a non-exportable secret in the chip's fuses that allows them to authenticate processors. Once a chip is authenticated and its capabilities checked, low-level primitives can be trusted to correctly validate hardware and software components of the TEE implementation. As a result of this process, the processor's manufacturer issues a \emph{platform certificate} (\PC{}) - we describe this certificate and its elements as follows.

\begin{itemize}[leftmargin=*]
	\item The manufacturer's public and private \emph{root of trust} keys \RoT{}$_{\pb{}}$ and \RoT{}$_{\pv{}}$, respectively, are elements of a signature scheme used to authenticate messages issued by this actor. The public component is well-known and trusted to be the correct \emph{root of trust} key for the manufacturer.
	
	\item The platform's public and private \emph{protected} keys \Plat{}$_{\pb{}}$ and \Plat{}$_{\pv{}}$, respectively, are elements of a signature scheme used to authenticate messages issued by this actor, and more specifically its TEE. These keys are managed by the TEE implementation and not even the platform owner has access to the private components.
	
	\item $\PC{} = (\Plat_{\pb{}})_{\RoT{}_{\pv{}}}$ is a certificate containing the platform's protected public key and a cryptographic signature of this key issued by the processor's manufacturer using \RoT{}$_{\pv{}}$. 
	
	\begin{itemize}
			\item In this paper, we use  $(el_1,\ldots,el_n)_{k}$ to denote a certificate containing the elements $el_1,\ldots,el_n$ together with a cryptographic signature of them using key $k$.
	\end{itemize}
\end{itemize}

This certificate is evidence that the platform's TEE is valid and has been properly set up, and its authenticity can be verified using \RoT{}$_{\pb{}}$. It also vouches for the platform protected key $\Plat_{\pb{}}$. As we detail later, this certificate and the elements above play a part in the attestation process.

As shown in Figure~\ref{fig:protocol}, the protocol works in three phases, \ie, initialization, 
remote attestation and program analysis. We now describe them in detail.

\subsection{Initialization}
\label{subsec:initialization}

The initialization phase of \tcpa begins with Producer and Consumer agreeing on the program analysis framework $X$, building framework $B$, and regulatory properties $P$ that are to be used by our isolated computation \TCPAIC{}. Once they have agreed on these parameters, Bob creates the isolated computation $\TCPAIC{}(X,B,P)$ as he is the owner of the platform. Thus, the code and data used by $\TCPAIC{}(X,B,P)$ is loaded in encrypted form, measured, and executed. Once this service starts, it generates its own key pair using a trusted source of randomness, and the \emph{isolated computation certificate} ($\ICC$). We define these elements as follows. 

\begin{itemize}[leftmargin=*]
	\item The $\TCPAIC{}(X,B,P)$ public and private keys \IC{}$_{\pb{}}$ and \IC{}$_{\pv{}}$, respectively, are elements of a signature scheme used to authenticate messages issued by this actor. These keys are managed by the isolated computation process $\TCPAIC{}(X,B,P)$ and not even the platform owner has access to the private components.
	
	\item $\ICC{} = (m_{\IC{}}, \IC{}_{\pb})_{\Plat{}_{\pv{}}}$ is a certificate containing the cryptographic measurement of the isolated computation process $\TCPAIC{}(X,B,P)$ given by $m_{\IC{}}$, and $\TCPAIC{}(X,B,P)$'s public key $\IC{}_{\pb{}}$. The certificate is signed by the platform's protected private key $\Plat{}_{\pv{}}$. The measurement $m_{\IC{}}$ is a cryptographic hash of the memory pages, \ie, code and data, loaded into main memory corresponding to $\TCPAIC{}(X,B,P)$; it accounts for the parameters $X$, $B$ and $P$. 
\end{itemize} 

This certificate provides evidence that an isolated computation with measurement $m_{\IC{}}$ has been created by a platform identified by key $\Plat_{\pb{}}$, and that this computation certified the blob of data, which happens to be its public key, $\IC{}_{\pb{}}$. While the measurement of the computation is calculated and set by the TEE, the isolated computation is free to pass any extra blob of data to be certified. Thus, while the measurement can be trusted to be correct, provided that the platform and TEE are trusted, the blob of data must only be relied upon if the isolated computation's code request the certification of the appropriate data.

\subsection{Remote Attestation}
\label{subsec:attestation}

The remote attestation process should provide enough evidence to convince Alice (Provider) that the expected isolated computation has been properly set up and started in a valid TEE-enabled platform. This phase begins by Alice requesting both the platform and isolated computation certificates $\PC{}$ and $\ICC{}$. Once she receives them, Alice can verify this \emph{attestation certificate chain}. She knows and trusts the manufacturer's $\RoT{}_{\pb{}}$ key. Therefore, she can use this key to verify $\PC{}$'s signature/authenticity. Once this certificate is verified, she can extract the platform protected key $\Plat{}_{\pb{}}$ from it and use this key to verify $\ICC$. The last step in this verification process consists of checking for the expected measurement. Alice calculates the measurement that she expects for $\TCPAIC(X,B,P)$ - let us call is $m_{exp}$ - and then compares it to the $m_{\IC{}}$ element of $\ICC{}$. If any of these verifications fail, Alice stops taking part in the protocol. If all of them succeed, however, Alice should be convinced that $\TCPAIC(X,B,P)$ has been correctly isolated, it is the computation expected, and, as part of its execution, its public key $\IC{}_{\pb{}}$, an element of certificate \ICC{}, has been certified. In the final step in this phase Alice and $\TCPAIC(X,B,P)$ negotiate $T$: a symmetric encryption scheme key used to confidentially transmit data between them. This key is negotiated in a way that it does not become known to Bob, say using some form of the Diffie–Hellman key exchange method. The key pair $\IC_{\pb{}}$ and $\IC_{\pv{}}$ can be used to authenticate $\TCPAIC(X,B,P)$'s messages in this negotiation process. 


\subsection{Program Analysis}
\label{subsec:ssg}

\begin{figure}[t]
	\centering
	\includegraphics[width=\linewidth]{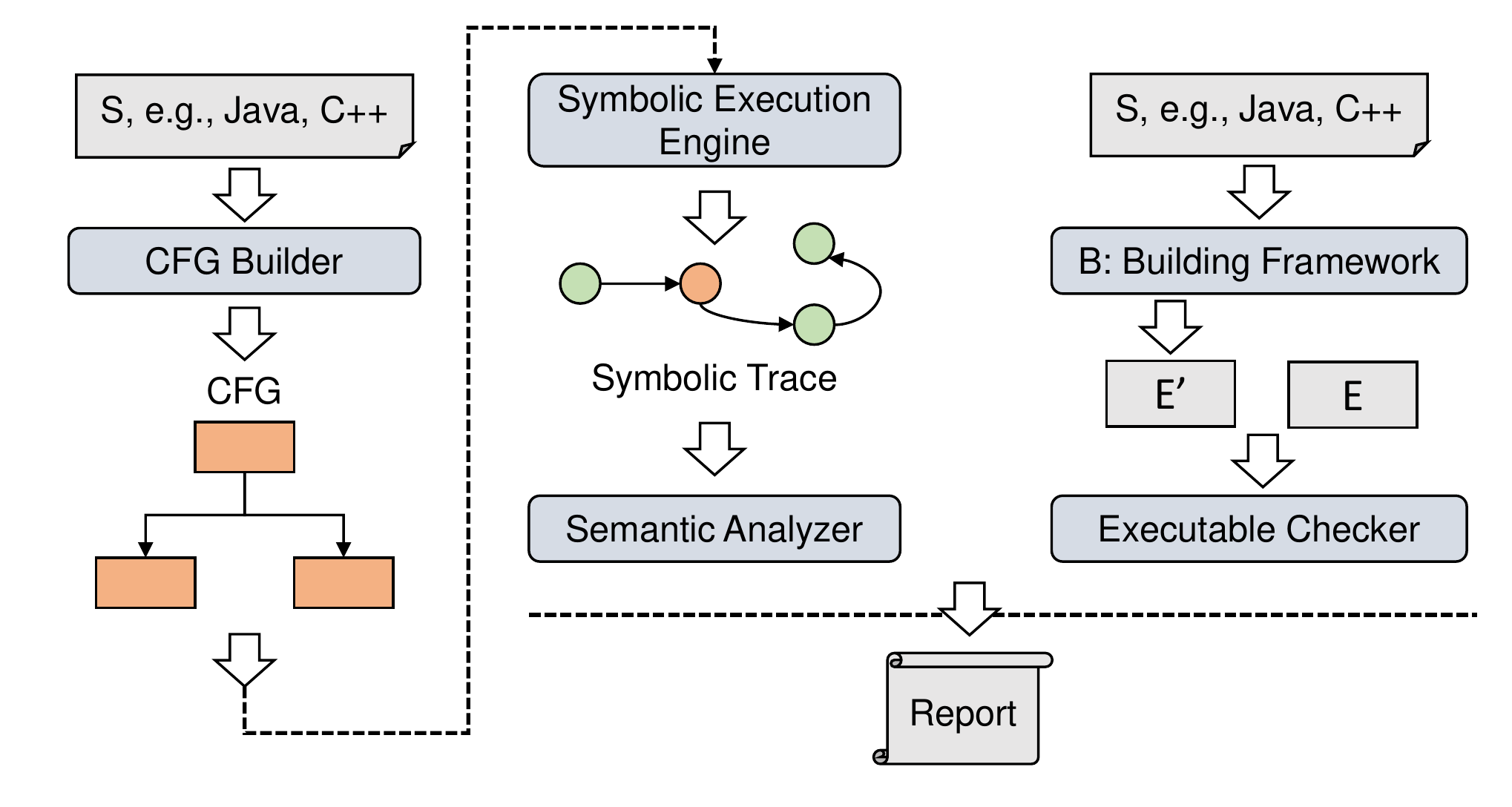}
	\caption{\label{fig:program-analysis}The workflow of program analysis.}	
\end{figure}

The program analysis phase of our protocol involves the examination of $S$ and $E$. 
Alice starts this phase by sending $S_T$ and $E$ to $\TCPAIC{}(X,B,P)$. 
The isolated computation, then, decrypts $S_T$ and proceeds to check: (i) whether $S$ meets the properties $P$, and (ii) whether $E$ is an executable capturing the behaviour of $S$. In principle, \tcpa does not limit the type of analysis 
run by $\TCPAIC{}$, that is, any well-defined framework could be used to verify $S$'s compliance with $P$ and $E$'s correspondence to $S$. Figure~\ref{fig:program-analysis} illustrates a framework to check both (i) and (ii).
The left part demonstrates a typical analysis flow based on symbolic execution. The source code 
of a program is firstly taken by \cfg builder to generate a control flow graph , \ie, \cfg. 
Then, a symbolic execution engine is triggered to explore the \cfg with symbolic inputs. 
In the process of symbolic execution, symbolic traces are produced for further analysis. 
For a specific trace, a semantic analyzer checks whether the properties hold or not. Lastly, 
an analysis report is generated as a summary of the process. The design 
of a specific program analysis technique is not our main focus in this paper, and as mentioned before, other verification frameworks could be plugged into our \tcpa protocol. The right part of this framework describes a process to determine $E$'s correspondence 
to $S$. Specifically, an executable $E'$ is built with $B$ for the given $S$. 
Then, $E'$ is compared with $E$ (\ie, executable provided by Alice) using an executable checker that syntactically compares them for equality. If so, it is safe to conclude that $E$ corresponds to $S$. Of course, we could have more sophisticated executable checkers where some form of \emph{semantic} equivalence between $E$ and $E'$ could be checked instead. Furthermore, we could even completely bypass this comparison step, and return $E'$ to Alice. Given that this executable was created by $\TCPAIC{}(X,B,P)$ using $B$ and that $B$ is trusted to be correct, $E'$ should correspond to $S$. Once the analysis is finished, $\TCPAIC{}(X,B,P)$ creates the \emph{compliance certificate} (\CC{}) as follows.

\begin{itemize}[leftmargin=*]
	\item $\CC{} = (h(S), E, R)_{\IC{}_{\pv{}}}$ contains a cryptographic hash of $S$ (i.e., $h(S)$), the executable $E$ sent by Alice, and an analysis report $R$, and it is signed using the computation $\TCPAIC{}(X,B,P)$'s private key $\IC{}_{\pv{}}$ . The report $R$ is a pair $(eo, (po_1,\ldots,po_{n}))$ where $eo$ is the executable equality outcome - a boolean that is true if they are equivalent, and false otherwise - and the tuple $(po_1,\ldots,po_{n})$ has a property verification outcome $po_i$ for each corresponding to property $p_i$ in $P$. The format of property outcomes $po_i$ will vary based on the choice of source code and property languages, and verification framework. For instance, considering stochastic systems and verification frameworks, a property outcome would present the likelihood of a property being valid. On the other hand, a non-stochastic model checker applied to a deterministic program could output precisely whether a property holds. There are even non-stochastic approximate analysers which might admit an \emph{unknown} outcome when they cannot issue a definitive statement about the specific $S$ and $p_i$ being analysed. These approximations are typically used for the sake of decidability or efficiency.
\end{itemize}

Finally, $\CC{}$ is sent back to Alice which can forward $\PC{}$, $\ICC{}$ and $\CC{}$ to Bob. He can then verify this certificate chain just like Alice does on the remote attestation phase with the extra steps that $\CC{}$'s signature must be verified against $\IC_{\pb{}}$ and that the report $R$ in $\CC{}$ must have the executable equality outcome set to true and that all properties are valid. If the verification of this certificate chain succeeds, Bob trusts $R$ - and, consequently, the compliance status of $E$ - and can start using this product.

In this protocol, neither Alice or Bob authenticate any messages by, say, signing them. We could, of course, require that Alice and Bob have a key pair each, for which public component is known to the other participant, that they could use to set up authenticated and confidential channels between them using well-known cryptographic protocols. The protocol could then take place over these channels. We do not detail this set-up as our primary goal in this paper is to introduce a protocol by which Bob can be convinced of $E$'s (and $S$') compliance without having access to $S$ and not make it confidential to other parties.

Perhaps one interesting addition to the protocol would be a \emph{certificate of origin} provided by Alice. In the current version of the protocol, Bob can verify that $E$ meets the expected regulatory properties, but the protocol does not have any mechanism to tie $E$ and $S$ to Alice. Of course, Bob might be satisfied that $E$ complies with $P$ alone, regardless of who the author of $E$ or $S$ is. However, Bob could also demand to know their author so that he can later raise a dispute related to $E$ against its author, for instance. To satisfy this demand, the protocol could be extended with an extra step by which Alice generates an origin certificate (\OC{}) containing the $h(S)$ and $E$, and cryptographically signed by its private key $A_{\pv{}}$ - the public counterpart $A_{\pb{}}$ of which is known to Bob. This certificate attests Alice's authorship and can be verified by Bob using $A_{\pb{}}$. This certificate together with $h(S)$ in $\CC{}$ commits Alice to the source code used in \tcpa. Thus, she cannot claim a different source code was used by the protocol in a possible dispute resolution process.

Note that, as the platform owner/TEE host, Bob has unrestricted access to the messages transiting between Alice and $\TCPAIC{}(X,B,P)$, and to the platform and TEE running this isolated computation, and yet he is unable to learn anything about the source code $S$ apart from its compliance status with respect to properties $P$ and that it gives rise to executable $E$. He does not have access to the source code when sent by Alice to $\TCPAIC{}(X,B,P)$ as the negotiated symmetric key $T$ is not known to him, and the TEE managing the execution and isolation of $\TCPAIC{}(X,B,P)$ prevents him from accessing this secret when its compliance analysis is being carried out.

Regulators may also impose restrictions on the verification framework $X$ used to verify whether $S$ meets $P$. They could, for instance, have a list of approved frameworks, and expect only those to be used in the context of $\TCPAIC{}$. The measurement of $\TCPAIC{}(X,B,P)$ - which depends on $X$, $B$, and $P$ - can be used to show that an approved framework has been indeed used. A party using our protocol ought to be able to independently calculate the measurement of $\TCPAIC{}(X,B,P)$, given $X$, $B$, and $P$, for the sake of confirming it is interacting with the appropriate isolated computation in the process of remote attestation. Therefore, it can use this trusted measurement to test if a given framework $X'$ was used and, then, check whether $X'$ is in the list of approved frameworks. More sophisticated procedures could be devised to ensure a given verification framework meets the norms of a regulator, including having its code verified for some properties using perhaps its own separate $\TCPAIC{}$ instantiation for that.

It is worth mentioning that verification frameworks can be non-deterministic, that is, for some frameworks, two executions checking the same system and property can give rise to different verification outcomes. The parties relying on our protocol must keep this sort of behaviour in mind as it can be maliciously exploited. Let us say, for instance, that a verification framework in very rare occasions fail to report that a property is violated, returning that it was unable to assert the property instead, and let us assume that this sort of \emph{unknown} result is good enough to pass some regulation. This sort of behaviour could be explained, for instance, if we had a verifier $X'$ that has a constraint on the number of states it can examine, and so, for a given input system $S'$, a large randomly-selected portion of its state space can be analysed but not all of it. In such a case, a malicious party could re-analyse $S'$ using our protocol until $X'$ misses the small fraction of states of $S'$ that witness the violation. Thus, the consumer $B$ might then need to know that the \tcpa analysis is instigated by someone he trusts such as $B$ himself - as it is in the detailed protocol we have set out. 
Program analysis typically produces reports in addition to a yes/no/unknown result. In \tcpa, the form of a report, \eg, 
an exception leading to a counter example, may need to be withheld from the software customer because it gives away details of 
the confidential sources.

%% file: design.tex
\subsection{Architecture}
\label{subsec:architecture}
\begin{figure*}[h]
\centering
\includegraphics[width=.7\linewidth]{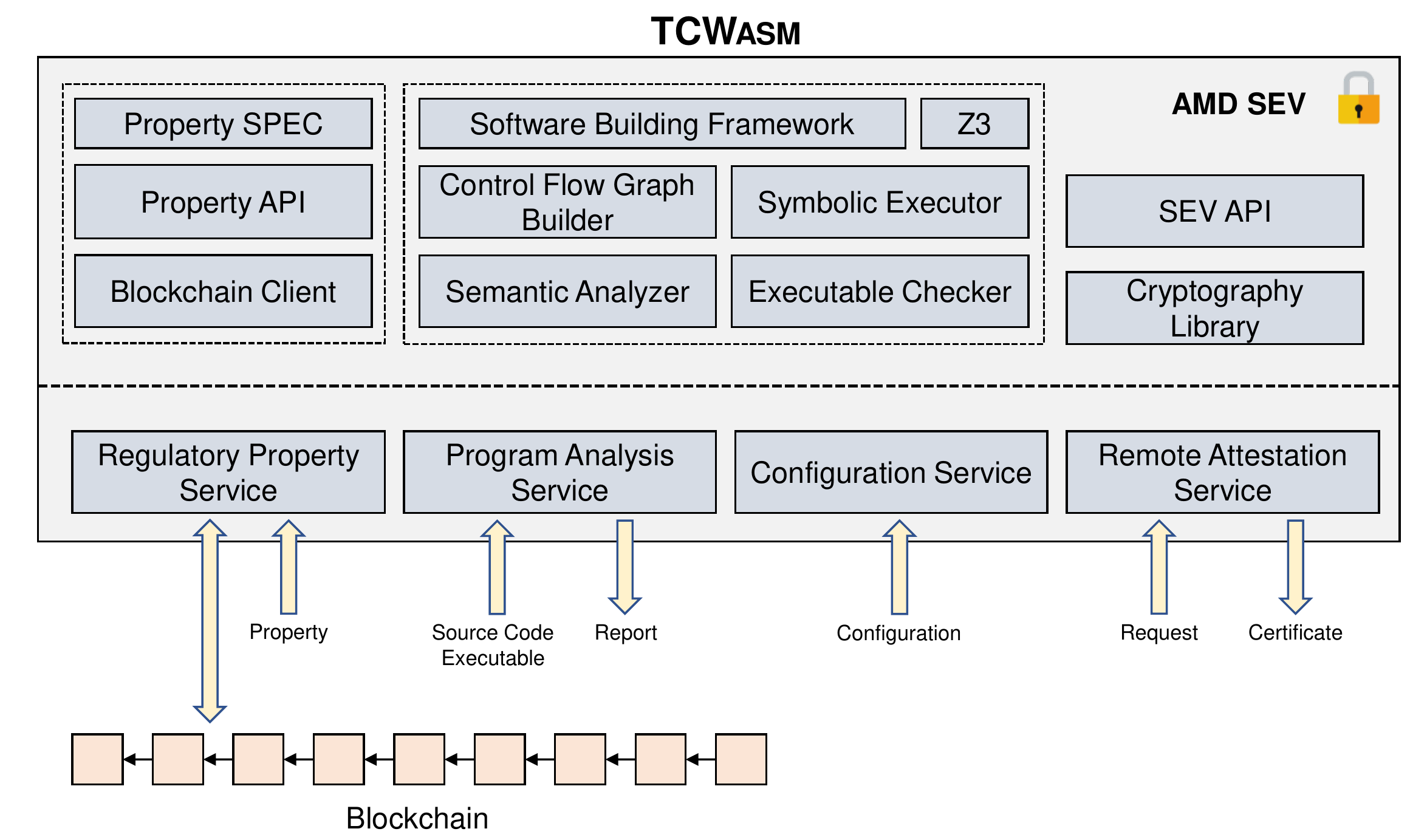}
\caption{\label{fig:sys}The architecture of \tool.}	
\end{figure*}

We realized the \tcpa protocol as described above in the \tool system to enable 
trusted and confidential program analysis for programs that compile to web assembly (\wasm). 
Specifically, we used the AMD SEV~\cite{sev2020strengthening} as the underlying trusted execution 
environment and a symbolic-execution-based engine to deliver program analysis~\cite{yang2020seraph}. 
The architecture of \tool is shown in Figure~\ref{fig:sys}. 

In general, \tool is implemented as a form of cloud service and is compatible with any existing 
platform, \eg, Amazon, Google, Alibaba etc. We offer four types of sub-services for 
end users, \eg, Alice and Bob in Figure~\ref{fig:protocol}, who are trying to agree on the 
regulatory compliance of a confidential software. The functionalities of these sub-services 
are summarized below.

\begin{itemize}[leftmargin=*]

\item \textbf{Regulatory Property Service} allows a software customer to publish an agreed set of 
regulatory properties which will be verified for the software created by a provider later in \tool. 
The agreement on properties is formed with signatures of both the provider and customer 
included. The service also offers accesses to properties that are publicly available. Any relevant 
participant - \eg, downstream users, audit teams, \textit{etc} - can check the properties managed on \tool 
and understand the compliance guarantee from their own perspective.

\item \textbf{Program Analysis Service} runs the underlying analysis processes to decide whether the 
given software is compliant with the specified set of properties. As illustrated in Figure~\ref{fig:protocol}, 
the provider is required to submit an encrypted version of source code and the deployed executable. 
The service will generate a certificate of compliance as a form of report for relevant parties, \eg, the 
customer, to confirm that the specified computation process has been executed on the given software and 
yielded a compliance decision for it to explain whether the software satisfies the properties, and why it 
is the case.

\item \textbf{Configuration Service} helps users configure \tool based on their preference. 
Practically, it is reasonable to make the projection that \tool will probably have a collection 
of program analysis settings instead of a single one. For example, different customers might like to adopt 
different algorithms to analyze the software. The configuration service allows users to specify the computation 
without looking into too many technical details. 

\item \textbf{Remote Attestation Service} enables an end user to verify the running environment 
on the cloud. To this end, a user is required to post a request to \tool and start the remote attestation 
process. The service sends back a cryptographic certificate to guarantee that the running environment is indeed 
as stated (\eg, an AMD SEV) and the image to be loaded and executed is as expected. The certificate can be passed to 
other parties and they can then confirm the validity of the running environment by checking the certificate. 
The remote attestation service can be contacted at any time during the execution to generate the certificate as long 
as it is necessary to do so.

\end{itemize}

\subsection{Trusted Environment}
\label{subsec:env}

We use AMD SEV as the underlying \tee environment for \tool. A cryptography library is included to realize 
encryption and decryption actions as specified in \tcpa based on the set of cryptographic keys provided in AMD SEV. 
In addition, \tool implements a set of APIs to deliver the trusted certificate from AMD SEV to end users, \eg, a 
verifiable measurement of firmware, disk, memory, \textit{etc}. Due to the portability of AMD SEV, integration with other 
program analysis engines is straightforward.

\subsection{Property Specification and Management}
\label{subsec:property}
In \tool, the agreed set of regulatory properties is stored on a blockchain. Therefore, a majority of nodes 
in the network will have a consistent view on those properties as they replicate the ledger and states. To 
this end, \tool embeds a blockchain client to interact with the underlying blockchain. In principle, both 
permissioned and permissionless blockchains can fit in our setting, \eg, \ethereum~\cite{wood2014ethereum}, 
Fabric~\cite{cachin2016architecture}, \textit{etc}. While the former is more suitable for a relatively stable group of 
software providers and customers, the latter fits better in an open market where any global participant is 
allowed to join. The blockchain client can submit properties provided by users to blockchain in several different 
ways, \eg, as transaction metadata, as structural storage of smart contracts, \textit{etc}. In cases of dispute (\ie, 
providers and consumers have conflicting views on properties) which is commonly resulted from forking on the 
blockchain, \tool is able to have a further decision after several blocks and the longest chain can be identified. 

Moreover, properties can be defined with both existing or user-defined specification languages, \eg, computation 
tree logic, linear temporal logic, \etc In the preliminary realization of \tcpa, we specify safety properties as 
program assertions, \eg, via \texttt{assert} statement. Therefore, the checking of such properties is converted to 
a reachability problem where an analysis is required to search for a transition that can potentially lead to a 
unsafe state. As mentioned earlier, it is possible to have multiple types of specifications in \tool. Users are allowed to configure 
which one they will apply for the submitted properties.


\subsection{A Formal Framework For Analyzing \wasm}
\label{subsec:formal}
We formalize a \wasm program $P \coloneqq \langle \mathcal{M}_0, \mathcal{M}_1, \cdots, \mathcal{M}_n \rangle$ as a group of modules to be deployed, each of 
which is a tuple $\mathcal{M} \coloneqq \langle \mathbb{T}, F, T, M, G \rangle$, where 
\begin{itemize}[leftmargin=*]
\item $\mathbb{T}$ defines a vector of function types.

\item $F \coloneqq \langle F_0, F_1, \cdots, F_n \rangle$ is a collection of functions. For each function $f\in F$, 
$ f = \langle \textit{ft} \in \mathbb{T}, L, E\rangle$ comes with a type, a set $L$ of local variables 
that are only accessible inside a function and a list $E$ of expressions. More specifically, expressions in $E$ are 
converted to a control flow graph, where a node in the graph is a basic block that includes a sequence of 
\wasm statements and an edge is a jump from one basic block to another.
 
\item $T$ is a table of indexed function references.

\item $M$ is a list of memory elements which can be linearly addressed inside a module.

\item $G$ is a set of global variables which can be assessed from any function.
\end{itemize}

The program analysis framework $A$ that we developed in this work for \wasm programs is formalized 
as a tuple $A \coloneqq \langle K_f, K_o, M, \mathit{SrcMap}, \mathit{Spec} \rangle$, 
where $K_f$ and $K_o$ are stacks of function calls and operands, respectively. 
$M$ denotes the set of modules to be analyzed by $A$. $\mathit{SrcMap}$ is a source map 
which creates a bidirectional link between confidential source code and bytecode of \wasm. 
$\mathit{Spec}$ is the specification standard of \wasm. Moreover, a configuration 
of $A$ is abstracted as $c \coloneqq \langle \mathit{pc}, \mathit{src}, k_f, k_o, S, P, \mathcal{G} \rangle$, 
where 
\begin{itemize}[leftmargin=*]
\item $\mathit{pc}$ is a program counter that points to the current program location.

\item $\mathit{src}$ is a source code pointer which maps from current $\mathit{pc}$ to 
the corresponding source code (\eg, at a specific line of code and offset).

\item $k_f$ shows the current stack of functions.

\item $k_o$ describes the current stack of operands.

\item $S \coloneqq \langle S_0, S_1, \cdots, S_n \rangle$ is a tuple containing one state per module. 
Specifically, $S_i \in S$ ($0\le i \le n$) is the state for module $\mathcal{M}_i$, which is a 
tuple $\langle m, G, L\rangle$. $m$ is a snapshot of memory. $G$ and $L$ describe valuation for 
both global and local variables. 

\item $P$ is a set of path conditions, each of which is a group of symbolic conditions to enable 
the exploration of a specific program path.

\item $\mathcal{G} = \langle V, E \rangle$ is a semantic graph that describes fundamental program 
relations. The vertices $V$ represent variables, \eg, local variable, global variable, \etc Edges 
$E$ are an abstraction of dependency between variables. In general, we defined two types of dependencies, 
\ie, data flow and control flow dependency (marked as $\mapsto$ and $\rightarrowtail$), respectively. 
For example, $a \mapsto b$ indicates that the value of $b$ is dependent on the value of $a$ (data flow). 
$c \rightarrowtail d$ defines a control flow relation where the value of $c$ is related to the path selection 
of a control flow therefore affects the value of $d$.
\end{itemize}

As introduced above, the current implementation of program analysis in \tcpa is based on 
symbolic execution. Specifically, the configuration of $A$ is initialized with a starting 
module $M_j$ and function $f_k$. The analysis start by examining the statements in this function's body.
Given a statement $e$, the program analysis framework $A$ retrieves elements from $K_o$, performs 
computation as specified in the \wasm standard, potentially updating memory, local storage, 
global storage, path conditions and semantic graph. That said, the processing of $e$ yields a transition 
of configuration of $A$, from the current one $c \coloneqq \langle \mathit{pc}, \mathit{src}, k_f, k_o, S, P, \mathcal{G} \rangle$ 
to a new one $c'$. When a program path has been visited, we combine the path conditions with the specified 
properties for SMT solving and further verifying whether the properties hold or not. This process iterates 
until a completion criterion for program analysis is realized, \eg, coverage, time, \etc

Although symbolic execution is performed on \wasm bytecode, it requires fundamental knowledge from the 
confidential source code. The most obvious information required in \tcpa is a source map that links 
every bit in the bytecode back to source code, and the other way around as well. With a source map, \tcpa 
can provide software producers with understandable results, \eg, a bug is located at a specific line. 
Moreover, type information at source code level is also important to facilitate program analysis. For example, we 
would be able to associate memory access with variables in source code with their types.

The analysis framework $A$ can have different implementations. As Figure~\ref{fig:sys} shows, we currently use 
a framework based on symbolic execution but the general design can fit into a wider range of program analysis techniques, 
\eg, formal verification, automated testing, \etc

%% file: eval.tex
\subsection{Benchmark and Evaluation Setup}
\label{subsec:bench}

\myparagraph{Benchmark}
We have conducted a preliminary evaluation of \tool. The benchmarks used in the 
evaluation are shown in Table~\ref{tab:benchmark}. We used 8 examples 
from different application domains with \neval web assembly files in total. 
The size of benchmark files varies from 2KB to 840KB. We also counted the number of 
\wasm instructions in all files, which manifested a relatively large difference across 
different test cases. The largest one (Zxing) has over 381K instructions, while the smallest 
(Snake) only includes 528 instructions. The benchmarks and results are publicly available at \dataset.

\input{results}

\myparagraph{Evaluation Setup}
All the experiments were performed on two environments, \ie, with and without \tee, respectively. 
The \tee environment was set up on Google Cloud confidential computing 
platform\footnote{https://cloud.google.com/confidential-computing}, which uses 
the AMD Epyc processor and SEV-ES\footnote{https://developer.amd.com/sev/} as the underlying TEE setting. 
The cloud machine was configured with dual 2.25GHz cores, 8GB memory and a Ubuntu 18.04 
operating system. Moreover, the non-\tee environment carried Intel i9 processor with dual 2.3GHz 
cores, 8GB memory and a Ubuntu 18.04 operating system. All the computation was executed by only 
one core of the processor in our evaluation.

\subsection{Evaluation Results}
\label{subsec:results}
In the evaluation, we ran \tool with and without the support of TEE (\ie, AMD SEV 
in our case) on all benchmarks. For a test file $f$, \tool performs a systematic 
program analysis based on symbolic execution~\cite{king1976symbolic} to explore the 
state space of $f$ and create a semantic abstraction as well. That said, the execution 
did not include detection of specific types of bugs or errors, as in many existing program 
analyzers. The goal of this evaluation was to understand the performance trade-off with the 
design of \tcpa, rather than assessing the effectiveness of a certain bug-detection algorithm.
With a framework such as \tool, the implementation of a detector is a straightforward task 
even in the context of \tcpa.

\input{result-time}

\myparagraph{Time Overhead}
The time cost with and without a TEE is described in Table~\ref{tab:time}. 
In the evaluation, we observed the smallest 7.7\% overhead in the case of \texttt{tfjs-backend}, 
while the largest case was 367.5\% for \texttt{imagequant}. The average time overhead was 
139.3\% on all benchmark files. For a subset of the test cases, files with larger sizes introduced 
bigger time overhead as expected. For instance, \texttt{binjgb}, \texttt{module1}, \texttt{avif\_dec} 
led to an increasing level of overhead with a growing size and number of instructions. However, 
there were exceptions in the evaluation where big files manifested small overheads. For example, 
in the case of \texttt{mozjpeg\_enc} (217 KB), running \tool is 95.1\% slower than the non-TEE version. 
For the case of \texttt{rotate} (14 KB) which is only 6.5\% as large as \texttt{mozjpeg\_enc}, the 
overhead was 107.0\% that amounts to a relative 12.5\% growth. Further discussions on root causes of 
the overhead can be found below.


\input{result-memory}

\myparagraph{Memory Overhead}
In addition to time cost, we analyzed the memory overhead with \tool in our evaluation. 
Similarly, the analysis was conducted with and without the support of TEE, 
as shown in Table~\ref{tab:memory}. In general, majority of the overheads were below 45\%. 
More specifically, the overheads for half of the test cases were even less than 8\%, which 
we believe is highly acceptable in practical scenarios. On the other hand, there were two cases 
that manifested a 2X and 4X overheads, although the actual memory used were not big, \ie, 10.8MB 
and 36.4MB respectively. The \texttt{tfjs-backend} file was particularly interesting due to the 
fact that \tool consumed less memory with a TEE than the non-TEE version. Detailed explanations 
are given in the following section.

\subsection{Discussions}
\label{subsec:discuss}
We now describe a further discussion on the empirical results with \tool to help understand its 
performance manifested in the evaluation. 

\myparagraph{General Explanation}
First of all, the runtime overheads in general introduced by \tool in our evaluation are easy to understand. 
In the case of time overhead, the execution of \tcpa protocol was encapsulated in a \tee environment, 
which encrypts and decrypts memory accessed by the running program, \ie, in our case the \tool implementation, 
therefore should last longer than running \tcpa without a \tee (139.3\% as described in Table~\ref{tab:time}), 
depending on how efficient the \tee is realized. 
On the other hand, \tool did not manifest a higher level of memory consumption than a non-\tee implementation 
for the majority of test cases used in the evaluation as shown in Table~\ref{tab:memory}, due to the fact that 
encryption and decryption of memory in a \tee are not memory-intensive procedures thus commonly require little 
extra memory in running \tcpa.

\begin{figure}[h]
\centering
\includegraphics[width=.85\linewidth]{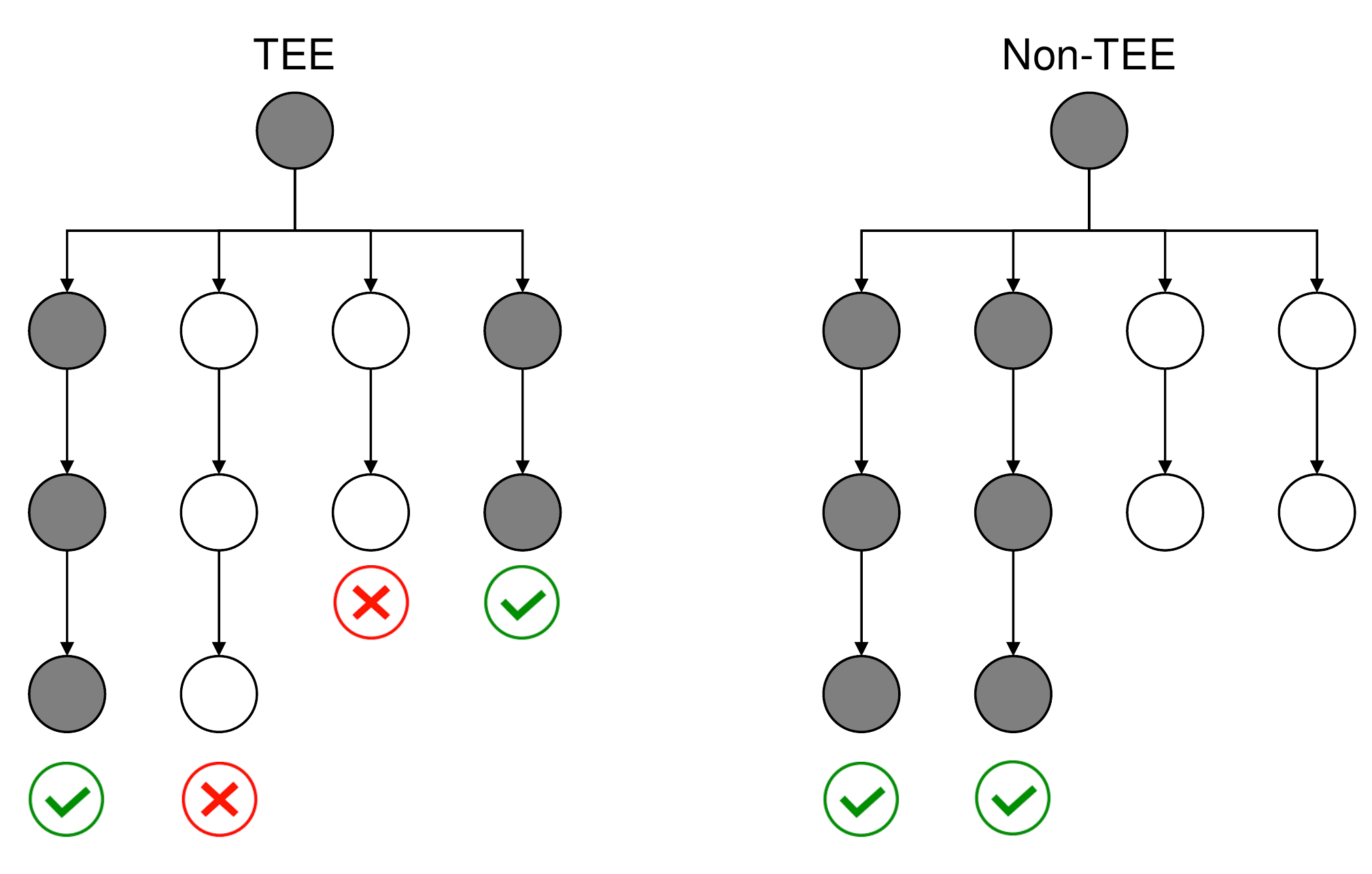}
\caption{\label{fig:setting}An evaluation setting of covering two paths with a specified timeout on each path.}	
\end{figure}

\myparagraph{Special Cases}
Despite the general analysis of evaluation results, we did observe that there were exceptions that seemed not to
be consistent with other cases. As shown in Table~\ref{tab:time}, it was much slower for \tool to process 
\texttt{avif\_dec}, \texttt{imagequant} and \texttt{asm-dom} than other test files. The average time overhead for 
the three is 273.0\% which almost doubles the number of total average. As explained above, the time overhead is 
mainly resulted from encryption and decryption of memory used by \tool. More specifically, the overhead is closely 
correlated to the memory complexity of analysis (\eg, the amount of memory used and the frequency to access it) 
adopted in the \tcpa realization, \ie, a symbolic-execution based analysis. Like many other well-designed symbolic 
engines, \tool uses a variety of specific data structures to store intermediate information of program analysis, 
\eg, states of analysis, symbolic contexts, path conditions, \etc Particularly, \tool introduced a graph-based structure to 
separate the modeling of a given program and its symbolic execution process. While the advantage of such design is 
to have better composability via integration with different symbolic execution engines and backend analyzers, it 
inevitably increases the level of memory consumption and access frequency. Moreover, abnormal time overheads were 
partially attributed to the evaluation setting as well. We use an illustrative example in Figure~\ref{fig:setting} 
to explain the cause. 

Figure~\ref{fig:setting} describes an evaluation setting to cover two paths in the given program and the exploration 
of each path is bounded within a specified timeout, \eg, one second. Although settings may vary across different 
program analyzers to deal with specific use cases, they commonly share similar fundamental parameters, \eg, level of 
coverage and timeout for SMT solving. In particular, Figure~\ref{fig:setting} demonstrates a scenario where setting 
overhead is introduced. Specifically, a program analyzer without \tee (left) manages to cover the first and second 
paths of a given program and then finishes the analysis without exploring the remaining paths. However, in the case 
on the right, the program analyzer with \tee (right) manages to cover the first path of a given program, fail at 
the second and third due to timeouts of SMT solving, and then cover the last. In such cases, although overheads 
on two visited paths are relatively small, the total overhead becomes much bigger because of unfinished explorations 
on the other two paths.

\input{result-validate}

In terms of memory overhead, the evaluation manifested abnormal results as well. Specifically, \texttt{imagequant} 
and \texttt{zxing} introduced a large overhead while \texttt{tfjs-backend} even showed a negative overhead, \ie, 
\tool was faster than the non-\tee version. Commonly, \tcpa does not introduce a high level of memory overhead because 
the encryption process, \eg, AES as used in our case with AMD SEV, often generates ciphertexts with similar sizes as 
plaintexts. However, there might be cases as well where ciphertexts are bigger with specific padding strategies. Another 
factor to potentially affect the measurement of memory overhead is garbage collection in virtual machines. For cases where 
memory consumption is measured right after a garbage collection process, we might have a much smaller number than expected. 
Further investigation on such cases is left as future work.

\myparagraph{Preliminary Validation}
Since the implementation of \tool is non-trivial, we conducted a preliminary validation with a small group of 
simple test cases to justify the root cause analysis as described above. The validation is shown in Table~\ref{tab:validate}.

Specifically, the test cases used in the validation included the following programs:
\begin{itemize}[leftmargin=*]
\item \texttt{self\_addition}: increment a variable $10^7$ times with a specific value

\item \texttt{array\_addition}: add to $10^7$ elements in a given array

\item \texttt{quick\_sort}: quick sort a given list

\item \texttt{constraint\_addition}: a program with an addition constraint for SMT solving

\item \texttt{constraint\_division}: a program with an division constraint for SMT solving
\end{itemize}

As shown in the first two rows of Table~\ref{tab:validate}, \texttt{self\_addition} manifested 
a relatively lower level of time overhead (13.6\%) compared to \texttt{array\_addition} (350.0\%). The gap 
is resulted from different structures of memory accessed by both programs. While 
\texttt{self\_addition} only manipulated a single unit of memory, \texttt{array\_addition} 
is allocated with a consecutive memory space therefore each access to it requires addressing 
with the starting point and the offset. As a result, running \texttt{array\_addition} with \tee 
was much slower than without \tee due to encryption of a more complicated memory. In the case 
of \texttt{self\_addition}, \tee did not slow down too much of the execution. Furthermore, a similar 
explanation can apply to \texttt{quick\_sort}, \ie, the third row of Table~\ref{tab:validate}. Since 
the memory used by a quick sorting algorithm commonly includes a pivot and sub-lists of a given list, 
it introduced a high level of time overhead (314.8\%) as \texttt{array\_addition}. Moreover, the last 
two rows of Table~\ref{tab:validate} demonstrated two cases with simple and complicated path 
constraints for SMT solving, respectively. While \texttt{constraint\_addition} generated a 
constraint with addition, \texttt{constraint\_division} was a division constraint. Therefore, it took 
longer for \tool to solve \texttt{constraint\_division} than \texttt{constraint\_addition}. As 
explained in Figure~\ref{fig:setting}, \tool managed to solve the division constraint without \tee 
but failed with \tee due to timeout (which could be verified based on runtime logs). 
Therefore, the time overhead in the forth row is larger, \ie, 92.3\%. On the 
other hand, the addition constraint can be solved with and without \tee thus did not manifest a large 
overhead in the last row. In terms of memory, all cases introduced slight overheads, which can be 
explained by the fact that the memory encryption process enforced by \tee (\ie, AMD SEV) 
did not require much extra memory space.

%% file: results.tex
\begin{table*}[htbp]
\centering
\caption{Statistics of the benchmark. KB: kilobyte. }\label{tab:benchmark}
\begin{tabular}{c | c | c | r | r}
\toprule
\textbf{Type} & \textbf{Benchmark} & \textbf{File} & \textbf{Size (KB)} & \#\textbf{Instruction} \\
\midrule
\multirow{6}{*}{Image Processing} & \multirow{5}{*}{Squoosh} & module1 & 224 & 111,533 \\

& & avif\_dec & 640 & 152,481\\

& & imagequant & 58 & 26,641\\

& & mozjpeg\_enc & 217 & 64,426\\

& & rotate & 14 & 700\\


& Zxing & zxing & 840 & 381,368\\

\hline

Machine Learning & Tensorflow & tfjs-backend & 222 & 108,735\\

\hline

\multirow{3}{*}{Library} & \multirow{3}{*}{C Standard Library} & stdio & 11 & 5,156\\

 &  & string & 11 & 4,664 \\

 &  & memory & 16 & 8,519 \\
 
\hline

Framework & asm-dom & asm-dom & 100 & 44,828\\

\hline

\multirow{3}{*}{Game} & Game Boy & binjgb & 98 & 41604\\

& Maze & maze & 2 & 780 \\

& Snake & snake & 2 & 528\\

\bottomrule
\end{tabular}
\end{table*}

%% file: result-time.tex
\begin{table}[htbp]
\centering
\caption{Time cost. s: second.}\label{tab:time}
\begin{tabular}{c r r r}
\toprule
\textbf{File} & \textbf{TEE (s)} & \textbf{Non-TEE (s)} & \textbf{Overhead} \\
\midrule

module1 & 67.7 & 36.3 & 85.4\% \\

avif\_dec & 1,823.6 & 559.6 & 225.9\% \\

imagequant & 493.0 & 105.4 & 367.5\% \\

mozjpeg\_enc & 659.7 & 338.2 & 95.1\% \\

rotate & 358.4 & 173.1 & 107.0\% \\

zxing & 1,427.5 & 530.4 & 169.3\% \\

tfjs-backend & 712.9 & 661.1 & 7.7\% \\

stdio & 53.5 & 20.4 & 162.3\% \\

string & 26.5 & 10.5 & 152.4\% \\

memory & 117.9 & 46.0 & 156.3\% \\

asm-dom & 382.5 & 117.5 & 225.5\% \\

binjgb & 155.5 & 136.7 & 13.8\% \\

maze & 2.2 & 1.0 & 120.0\%\\

snake & 0.5 & 0.3 & 61.3\%\\

\midrule
average & $\ast$ & $\ast$ & 139.3\% \\

%
%
%
%
%
%
%
%
%
%
%

\bottomrule
\end{tabular}
\end{table}

%% file: result-memory.tex
\begin{table}[htbp]
\centering
\caption{Memory cost. MB: megabyte.}\label{tab:memory}
\begin{tabular}{c r r r}
\toprule
\textbf{File} & \textbf{TEE (MB)} & \textbf{Non-TEE (MB)} & \textbf{Overhead} \\
\midrule

module1 & 6.4 & 5.7 & 12.3\% \\

avif\_dec & 80.2 & 59.5 & 34.8\% \\

imagequant & 51.2 & 10.8 & 374.1\% \\

mozjpeg\_enc & 78.5 & 75.9 & 3.4\% \\

rotate & 15.7 & 14.8 & 6.1\% \\

zxing & 106.0 & 36.4 & 191.2\% \\

tfjs-backend & 65.4 & 65.6 & -0.3\% \\

stdio & 11.7 & 10.9 & 7.3\% \\

string & 6.1 & 5.7 & 7.0\% \\

memory & 15.2 & 14.2 & 7.0\% \\

asm-dom & 52.3 & 49.4 & 5.9\% \\

binjgb & 20.6 & 14.2 & 45.1 \% \\

maze & 2.8 & 2.0 & 40.0\% \\

snake & 0.8 & 0.6 & 33.3\% \\

\midrule
average & $\ast$ & $\ast$ & 54.8\% \\

\bottomrule
\end{tabular}
\end{table}

%% file: result-validate.tex
\begin{table*}[htbp]
\centering
\caption{The preliminary performance validation with simple programs.}\label{tab:validate}
\begin{tabular}{c r r r r r r}
\toprule
\multirow{2}{*}{\textbf{File}} & \multicolumn{3}{c}{\textbf{Time (second)}} & \multicolumn{3}{c}{\textbf{Memory (MB)}} \\

 & \textbf{TEE} & \textbf{Non-TEE} & \textbf{Overhead} & \textbf{TEE} & \textbf{Non-TEE} & \textbf{Overhead}\\
\midrule

self\_addition & 14.2 & 12.5 & 13.6\% & 0.7 & 0.5 & 40.0\%\\

array\_addition & 3.6 & 0.8 & 350.0\% & 389.3 & 389.2 & 0.03\% \\

quick\_sort & 134.4 & 32.4 & 314.8\% & 387.4 & 392.5 & -1.3\%\\

constraint\_addition & 0.4 & 0.3 & 33.3\% & 128.6 & 128.2 & 0.3\%\\

constraint\_division & 5.0 & 2.6 & 92.3\% & 300.1 & 302.8 & -0.9\%\\

\bottomrule
\end{tabular}
\end{table*}

%% file: rw.tex
\myparagraph{\tee-based Technology}
The capabilities of \tee{}s have been widely exploited to achieve security, confidentiality and 
simplicity in many application domains. In the design of secure systems, Baumann~\etal 
proposed the notion of shielded execution on cloud platforms~\cite{baumann2015shielding}. 
Their work addresses the dual challenges of executing unmodified legacy binaries 
and protecting them from a malicious host. Similar ideas were adopted in data-processing 
and delegation-based systems to achieve integrity and security without trusting the service 
providers~\cite{hunt2018ryoan,matetic2018delegatee,schneider2019secure}. 
Moreover, Tsai~\etal demonstrated that a 
fully-featured library operating system 
can deploy unmodified applications with the support of a \tee{}~\cite{tsai2017graphene}. 
Shen~\etal further introduced secure and efficient multitasking on top of library 
operating systems with Intel SGX~\cite{shen2020occlum}. 
In the area of blockchain and cryptocurrency, \tee{}s are often considered a tool to enable 
trusted and privacy-preserving transactions. Matetic~\etal leveraged SGX enclaves 
to protect privacy of bitcoin light clients~\cite{matetic2019bite}. Cheng~\etal designed 
a \tee{}-based blockchain that executes transactions with confidential input, 
output and states~\cite{cheng2019ekiden}. Other attempts included building asynchronous 
access~\cite{lind2019teechain}, allowing real-time cryptocurrency exchange~\cite{bentov2019tesseract}, 
resource-efficient mining~\cite{zhang2017rem} and so on. Moreover, \tee{}s have also been 
involved in a diverse collection of optimizations on existing software and hardware, 
\eg, databases~\cite{sun2021building,zhou2021veridb,eskandarian2017oblidb,priebe2018enclavedb}, 
network functionalities~\cite{herwig2020achieving,duan2019lightbox,poddar2018safebricks}, 
storage systems~\cite{bailleu2021avocado,bailleu2019speicher,krahn2018pesos,arasu2017concerto}. 
In addition to applications of \tee{}s, their design has also been the topic on recent papers proposing improvements
interoperability~\cite{feng2016scalable,weiser2019timber}, performance~\cite{li2021confidential}, and
resilience~\cite{bahmani2021cure,dessouky2020hybcache}.

\myparagraph{Program Analysis}
Program analysis has been a mainstream research direction in the programming language 
community for decades. Formal verification techniques were proposed to verify high-level 
programs (usually specified in formal modeling languages) against given specifications 
of target systems, \eg, safety, liveness, \etc 
Clarke~\etal introduced the technology of model checking to systematically 
explore the state space of a system and check whether important properties hold or
not~\cite{clarke2009model}. Hoare and Roscoe proposed Communicating Sequential 
Processes (CSP) as a fundamental formalism to model and verify concurrent 
systems~\cite{hoare1978communicating,roscoe1998theory}. Alur~\etal further 
introduced timed automata to handle timed systems with properties based on 
temporal logic~\cite{alur1994theory}. In addition to automatic techniques, theorem proving 
was designed to deliver rigorous verification with manual or semi-automatic 
proofs~\cite{coquand1986calculus,nipkow2002isabelle}. On the other hand, program 
analysis has also been applied in practical systems with low-level code, \eg, C++, 
Java, x86 binary, JVM bytecode, \etc Based on whether the process requires actually 
executing a program, the analysis is generally categorized into two classes, \ie, static 
and dynamic analysis. In the context of static program analysis, a variety of researches 
have been proposed to address fundamental challenges of programs, \eg, understanding 
semantics~\cite{cousot1977abstract}, memory modeling~\cite{flanagan2002extended}, 
interprocedual analysis~\cite{reps1995precise,horwitz1990interprocedural}, 
multithreading~\cite{engler2003racerx}, \etc In contrast, dynamically approaches check 
programs by instrumenting the code and analyzing it on the fly. Representative types 
of solutions include fuzzing~\cite{rawat2017vuzzer,chen2018angora,godefroid2008grammar}, 
predictive analysis~\cite{flanagan2009fasttrack,bond2010pacer,burckhardt2010randomized} 
and symbolic execution~\cite{cadar2008klee,godefroid2012sage,chipounov2011s2e}. 
In general, the \tcpa framework proposed in this paper is compatible with well-defined 
types of program analysis and the combination of them as well.

%% file: conclusion.tex
In this paper, we have highlighted the trusted and confidential program 
analysis (\tcpa) as a trustless technology to achieve agreed regulatory compliance 
on software among multiple parties without revealing sensitive information 
about it. We designed the protocol of \tcpa for trusted execution 
environments and developed \tool as the very first implementation of \tcpa. 
In our preliminary evaluation with \wasm benchmark files, \tool demonstrated the 
potential to handle complicated cases without incurring too much overhead. 
Further instantiation of \tcpa with new types of program analysis and trusted computing 
solutions is left for future work. We hope that with the rapid development of new TEEs, there will be convergence on a 
security model that serves \tcpa well. We have shown that on AMD SEV, it is already a serious proposition.